\documentclass[aps,floats,superscriptaddress,showpacs]{revtex4}
\usepackage{amsmath}

\usepackage{epsfig}
\usepackage{graphicx}
\usepackage{dcolumn}
\usepackage{bm}

\def\gapp{\lower.35em\hbox{$\stackrel{\textstyle>}{\sim}$}}
\def\lapp{\lower.35em\hbox{$\stackrel{\textstyle<}{\sim}$}}
\begin{document}
\bibliographystyle{apsrev}
\title{Pomeranchuk instability in doped graphene}
\author{Bel\'en Valenzuela }
\affiliation{Departamento de F\'isica de la Materia Condensada\\
Facultad de Ciencias, Universidad Aut\'onoma  de Madrid,\\
Cantoblanco, E-28049 Madrid, Spain.}
\author{Mar\'{\i}a A. H. Vozmediano}
\affiliation{Instituto de Ciencia de Materiales de Madrid,\\
CSIC, Cantoblanco, E-28049 Madrid, Spain.}
\date{\today}
\begin{abstract} The density of states of graphene has Van
Hove singularities that  can be reached by chemical doping and
have already been explored in photoemission experiments. We show
that in the presence of Coulomb interactions the system at the Van
Hove filling is likely to undergo a Pomeranchuk instability
breaking the lattice point group symmetry. In the presence of an
on--site Hubbard interaction the system is also unstable towards
ferromagnetism. We explore the competition of the two
instabilities and  build the phase  diagram. We also suggest that,
for doping levels where the trigonal warping is noticeable, the
Fermi liquid state in graphene can be stable up to zero
temperature avoiding the Kohn--Luttinger  mechanism and providing
an example of two dimensional Fermi liquid at zero temperature.
\end{abstract}
%

\maketitle

\section{Introduction}

The recent synthesis of single or few  layers of
graphite\cite{Netal05,Zetal05} has permitted to test the singular
transport properties predicted in early theoretical
studies\cite{S84,Hal88,GGV96,Khv01} and experiments\cite{Ketal03}.
The discovery of a substantial field effect \cite{Netal04} and the
expectations of ferromagnetic behavior\cite{Eetal03} has risen
great expectations to use graphene as a reasonable replacement of
nanotubes in electronic applications.

The Fermi level of graphene can be tuned over a wide energy range
by chemical doping \cite{MBetal07,BOetal07}, or by gating
\cite{Netal05,Zetal05}. The full dispersion relation of graphene
(Fig. 2) shows several regions of special interest. The low doping
region near half filling has been the object of much attention due
to the behavior of quasiparticles as massless Dirac electrons. The
Fermi surface at low filling (cusps of the dispersion relation)
begins being circular. For increasing values of the electron
density the trigonal distortion appears and, finally, several Van
Hove singularities (VHS) develop at energies the order of the
hopping parameter $E\sim 2.7 eV$. The possibility of finding
interesting physics around these densities have been put forward
recently in an experimental paper reporting on angle resolved
photoemission (ARPES) results \cite{MBetal07}. Chemical doping of
graphene with up to the VHS has been demonstrated in
\cite{MBetal07}. The VHS  are of particular interest in the
structure of graphene nanoribbons \cite{E07,NRC08} and nanotubes
where observation of  Van Hove singularities using scanning
tunneling microscopy have been reported \cite{Ketal99}. It is
known that around the Dirac point due to the vanishing  density of
states at the Fermi level, short range interactions such as an
on--site Hubbard term $U$ are irrelevant in the renormalization
group sense \cite{GGV94,GGV99} and should not give  rise to
instabilities at low energies or temperatures. The possibility to
stabilize a ferromagnetic \cite{PGC05b} or superconducting
\cite{GGV01,UC07,H07} phase near half filling is very  unlikely.
Alternatively, at  densities around the VHS the physics  is
dominated by the high density of states and should resemble the
one discussed before in the framework of the high-$T_c$
superconductors \cite{M97}. This is the physical situation to look
for ferromagnetic or superconductivity \cite{G08} in graphene --or
graphite--. In ref. \cite{GGV00} it was argued that the most
likely instabilities of a system whose Fermi surface has VHS and
no special nesting features are p--wave superconductivity and
ferromagnetism. A very interesting competing instability that may
 occur when the Fermi surface approaches singular points is a
redistribution of the electronic density that induces a
deformation of the Fermi surface breaking the symmetry of the
underlying lattice. This phenomenon called Pomeranchuk instability
 \cite{P58} has been found in the squared lattice at the VHS
filling \cite{HM00,VV01,G01,LCG08} and has played an important
role in the physics of the cuprate superconductors
\cite{YK00,YK00b} and in general in layered materials \cite{Y08}.
It has also been discussed in a more general context in \cite{V03}
where it was argued that opening of an anisotropic gap at the
Fermi surface is an alternative to cure the infrared singularities
giving rise to Pomeranchuk instabilities. More recently the
Pomeranchuk instability has been studied in relation with the
possible quantum critical points in strongly correlated systems
\cite{NC05}. Due to the special symmetry of the Fermi surface of
the honeycomb lattice, the pattern of symmetry breaking phases can
be richer than these of the square lattice.

In this paper we study the Van Hove filling in graphene modelled
with a single band  Hubbard model with on--site U and exchange
Coulomb interactions V. We perform a mean field calculation and
show that Pomeranchuk instability occurs very easily in the system
in the presence of the exchange interaction V. Adding the on--site
interaction U allows for ferromagnetic ground states. We study the
 coexistence or competition of the two and build a phase diagram
in  the (U, V) space of parameters. In section \ref{method} we
construct the exchange V and on--site U interactions in the
honeycomb lattice and describe the method of calculation and the
results obtained.
In the last section  we discuss  some points related with the
physics of graphene at high doping levels.  In the context of the
original Kohn--Luttinger instabilities \cite{KL65} that
establishes that all Fermi liquids will be unstable at low enough
temperatures, we suggest that graphene at intermediate fillings
can provide an example of stable metallic system at zero
temperature \cite{FKT04a,FKT04b,FKT04c}.  Finally we set the lines
for future developments.

\section{The model}
\label{method} The deformation of the Fermi surface by the
electronic interactions is a classical problem in condensed matter
dating back to founding work of Luttinger \cite{L60}. We model the
system by a single band Hubbard model in the honeycomb lattice
with an on--site interaction $U$ and a nearest neighbor Coulomb
interaction $V$, and perform a self--consistent calculation along
the lines of refs. \cite{VV01,RLG08}.

The Hubbard hamiltonian is
\begin{equation}
H=t\sum_{{\bf i};\sigma} c^+_{{\bf i}\sigma} c_{{\bf i}\sigma}
\;+\;U\sum_{\bf i} n_{{\bf i}\uparrow} n_{{\bf i}\downarrow}\;+
V\sum_{<{\bf i} {\bf j};\sigma \sigma'>}n_{{\bf i \sigma} }n_{{\bf j
\sigma'}}   \;\;. \label{ham-fs}
\end{equation}
\begin{figure}
\begin{center}
\includegraphics[width=5cm]{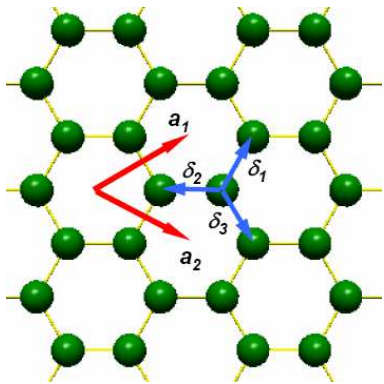}
\caption{(Color online) The honeycomb lattice of graphene.}
\label{lattice}
\end{center}
\end{figure}
The hopping parameter $t$ of graphene is $t\sim2.7eV$. The
next-to-nearest neighbor hopping $t'$ is estimated to be much
smaller  \cite{Oetal07} and it will be ignored. Due to the special
topology of the honeycomb lattice with two atoms per unit cell the
Hamiltonian is a $2\times 2$ matrix. We will work out in some
detail the different terms. The operators $\hat{a}, \hat{b}$
associated to the two triangular sublattices A, B are
\begin{equation}
a^+_{i A\sigma}=\frac{1}{\sqrt{N_A}}
\sum_k e^{i{\bf k. R_{i A}}}a^+_{{\bf k}\sigma} \;\;\;,\;\;\;
b^+_{i B\sigma}=\frac{1}{\sqrt{N_B}}
\sum_k e^{i{\bf k.R_{i B}}}b^+_{{\bf k}\sigma} \;\;\;,\;\;\;
\vec{R}_{i A}=\vec{R}_{i B}+\vec{\delta},
\end{equation}
where $N_A=N_B=N$ is the number of lattice cells and $\delta_i$
are the three vectors connecting a point of sublattice A with its
three neighbors in sublattice B (see Fig. \ref{lattice}).

In term of these operators the free Hamiltonian reads
\begin{equation}
H_0=t\sum_{k\sigma}[\phi({\bf k})a^+_{k\sigma}b_{k\sigma}
+\phi^*({\bf k})b^+_{k\sigma}a_{k\sigma}]
+\mu\sum_{k\sigma}(a^+_{k\sigma}a_{k\sigma}+b^+_{k\sigma}b_{k\sigma}),
\end{equation}
where we have included a chemical potential $\mu$.

\begin{figure}
\begin{center}
\includegraphics[width=8cm]{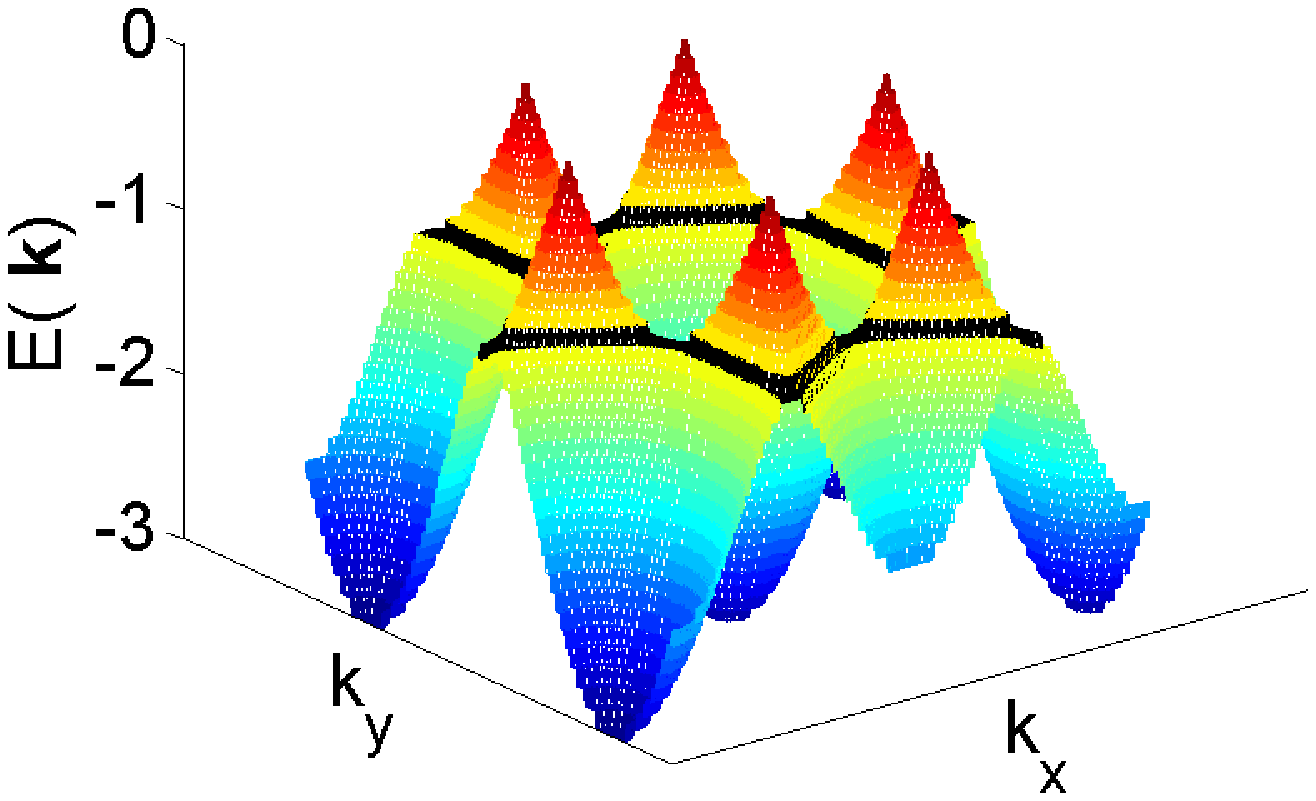}
\includegraphics[width=6cm]{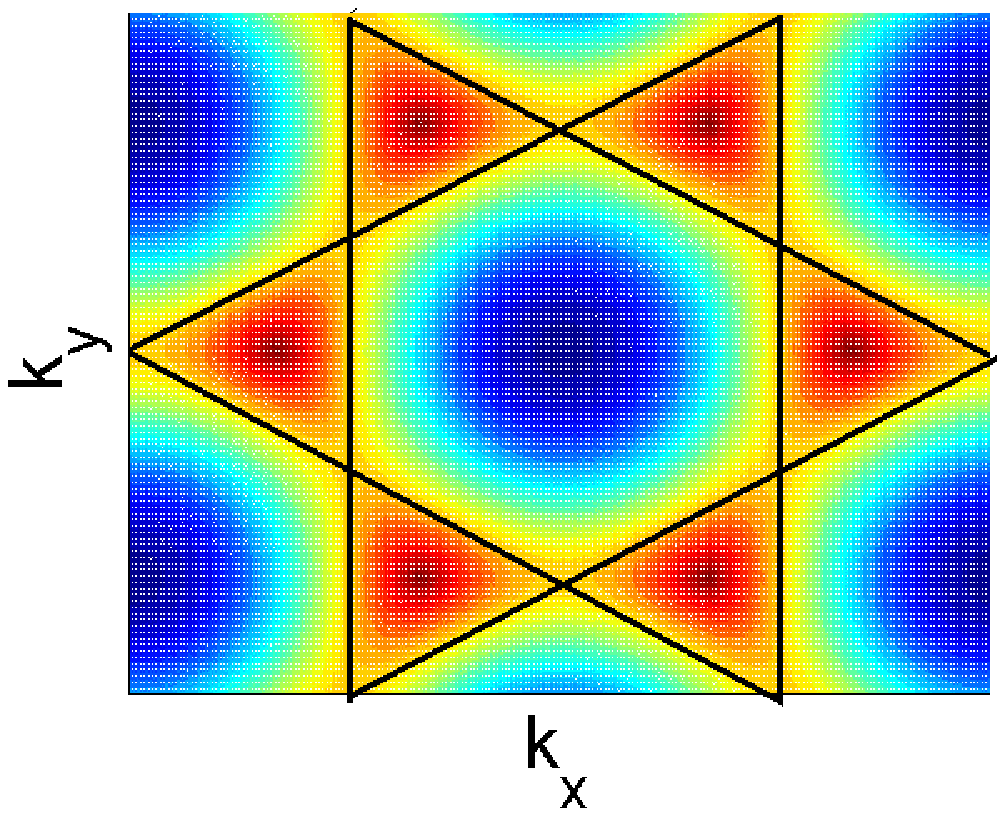}
\caption{(Color online) Left: Lower band of the dispersion
relation of graphene. The black line marks the Fermi surface for
the Van Hove filling. Right: Evolution of the Fermi surface of
graphene with doping. At half filling the Fermi surface consists
on the six vertices of the hexagonal Brillouin zone (black
points). Only two are independent, the rest being equivalent by
lattice vectors. Emptying the half filled band the Fermi surface
develops circular hole pockets that undergo the trigonal
distortion (in red) and become the yellow triangles at the Van
Hove fillig $\mu=t$. } \label{band2}
\end{center}
\end{figure}

In matrix form  the non--interacting Hamiltonian is
 \begin{eqnarray}
 H_0=\left(%
 \begin{array}{cc}
   \mu & t\phi({\bf k}) \\
  t\phi^*({\bf k}) & \mu\\
\end{array}%
\right), \label{H0}
\end{eqnarray}
where
\begin{equation}
\phi(\vec{k})=\sum_i e^{-i\vec{k} .\vec{\delta}_i},
\end{equation}
 The Hamiltonian (\ref{H0}) gives rise to the
dispersion relation
\begin{equation}
\varepsilon^0({\bf k})=\mu\pm t\sqrt{\vert\phi(\vec{k})\vert^2} =
\mu\pm t\sqrt{1+4\cos^2\frac{\sqrt{3}}{2}ak_x+
4\cos\frac{\sqrt{3}}{2}ak_x\cos\frac{3}{2}ak_y} \;,
\label{disprel}
\end{equation}
whose lower band is shown in the left hand side of Fig.
\ref{band2}.

\section{Coulomb interaction V and Pomeranchuk instability}

We will begin by studying the influence of the interaction $V$
omitting the spin of the electrons. The on-site U will be added to
study the competition of the Pomeranchuk instability with
ferromagnetism. The  interaction V takes place between nearest
neighbors belonging to opposite sublattices. It reads:
\begin{equation}
H_V=V\sum_{{\bf kk'q}}[a^+_{\bf k}a_{ {\bf k+q}}b^+_{{\bf
k'}}b_{{\bf k'-q}}\phi^*({\bf q})+b^+_{{\bf k}}b_{{\bf
k+q}}a^+_{{\bf k'}}a_{{\bf k'-q}}\phi({\bf q})].
\end{equation}
The mean field  Hamiltonian becomes
\begin{equation}
H_{MF}=\sum_{\bf k}[\tilde{\phi}({\bf k}) a^+_{{\bf k}}b_{{\bf
k}}+\tilde{\phi}^*({\bf k}) b^+_{{\bf k}}a_{\bf k}]+\mu\sum_{\bf
k}(a^+_{\bf k}a_{\bf k}+b^+_{\bf k}b_{\bf k}), \label{HMF}
\end{equation}
where
\begin{equation}
\tilde{\phi}({\bf k})=t\phi({\bf k})-\sum_{{\bf q}} V^*({\bf q})
<b^+_{{\bf k+q}}a_{{\bf k+q}}> \;\;\;,\;\;\;
V({\bf q})=\frac{V}{N}\sum_{{\bf q}}\phi^*({\bf q}),
\end{equation}
and $N$ is the number of lattice cells. In matrix form and
including the chemical potential the mean field Hamiltonian reads:
\begin{eqnarray}
H_{MF}=\left(%
\begin{array}{cc}
\mu & t\phi({\bf k}) -\sum_{\bf q}V^*({\bf q})<b^+_{k+q}a_{k+q}> \\
t\phi^*({\bf k})-\sum_{\bf q}V({\bf q})<a^+_{k+q}b_{k+q}>  & \mu\\
\end{array}
\right).
\label{HM}
\end{eqnarray}

The mean field Hamiltonian (\ref{HM}) gives rise to the dispersion
relation
\begin{equation}
E({\bf k})=\mu\pm[t^2\vert\phi({\bf k})\vert^2+F(V, {\bf k})]^{1/2},
\label{EV}
\end{equation}
where $F(V, {\bf k})$ is
\begin{eqnarray}
 F(V,{\bf k})&= -t V\sum_{{\bf q}} \phi^*({\bf
k})\phi({\bf q}) <a^+_{k+q}b_{k+q}> -t V \sum_{{\bf q}}\phi({\bf
k})\phi^*({\bf q})
 <b^+_{k+q}a_{k+q}> \\ \nonumber
&+
 V^2\sum_{{\bf q q'}}\phi^*({\bf q})\phi({\bf q'})
<a^+_{k+q}b_{k+q}><b^+_{k+q'}a_{k+q'}>.
\end{eqnarray}

We look for a self--consistent solution of eq. (\ref{EV}) imposing
the Luttinger theorem  {\it i. e.} that the area enclosed by the
interacting Fermi line is the same as the one chosen as the
initial condition. We begin with an initial (free) Fermi surface,
as an input,  add the interaction  and let it evolve until a
self--consistent solution is found with a given "final"
interacting Fermi surface. The important constraint is that the
total number of particles should remain fixed. This procedure has
been used in the same context in \cite{VV01,RLG08}.

\begin{figure}
\begin{center}
\includegraphics[width=5cm]{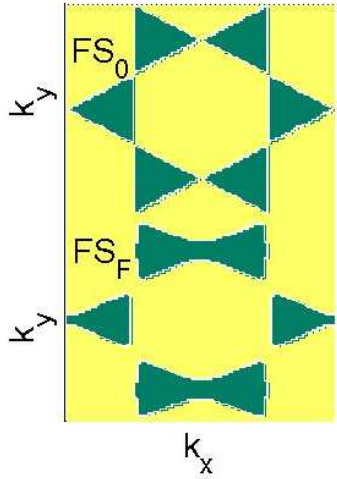}
\caption{Pomeranchuk instability suffered by an initial Fermi
surface at the Van Hove singularity (upper part) when an exchange
interaction  V=1t is added (lower part).}
\label{deformed}
\end{center}
\end{figure}
The Fermi surface of graphene around the Van Hove singularity is
shown in Fig. \ref{band2}. To better visualize the Pomeranchuk
instability we keep the image of the neighboring Brillouin zones.
Fig. \ref{deformed} shows the initial Fermi surface sitting at the
VHS (upper part) and the spontaneous deformation obtained
self--consistently from eq. (\ref{EV}) for a value of the exchange
interaction $V=t$.

\section{Ferromagnetism and competition of ferromagnetism and
Pomeranchuk instabilities} \label{FM}

In order to study  a possible ferromagnetic instability we will
add to the free Hamiltonian  an on--site Hubbard term
\begin{equation}
H_U=U\sum_i[a^+_{i\uparrow}a_{i\uparrow}a^+_{i\downarrow}a_{i\downarrow}+
b^+_{i\uparrow}b_{i\uparrow}b^+_{i\downarrow}b_{i\downarrow}],
\end{equation}
A mean field ferromagnetic state will be characterized by
\begin{equation}
a^+_{i\sigma}a_{i\sigma}=<a^+_{i\sigma}a_{i\sigma}>+\delta
(a^+_{i\sigma}a_{i\sigma})
\end{equation}
with
\begin{equation}
<a^+_{i\sigma}a_{i\sigma}>=\frac{n}{2}+\sigma\frac{m}{2}\;\;\;,\;\;\;
<b^+_{i\sigma}b_{i\sigma}>=\frac{n}{2}+\sigma\frac{m}{2}\;\;\;,\;\;\;\sigma=\pm,
\end{equation}
what produces the mean field Hamiltonian
\begin{equation}
H^U_{MF}=U\sum_{i\sigma}(\frac{n}{2}+\sigma\frac{m}{2})
(a^+_{i\sigma}a_{i\sigma}+b^+_{i\sigma}b_{i\sigma})-\frac{1}{2}U(n^2-m^2)N.
\end{equation}
Adding the kinetic term we get the  Hamiltonian
\begin{eqnarray}
H_0+H^U_{MF}=\left(%
\begin{array}{cc}
\mu +\frac{U}{2}(n+\sigma m)& t\phi({\bf k}) \\
t\phi^*({\bf k})  & \mu+\frac{U}{2}(n+\sigma m)\\
\end{array}
\right) \label{HUM}
\end{eqnarray}
whose dispersion relation
\begin{equation}
E_{k\sigma}=\mu_\sigma\pm t\vert\phi({\bf k})\vert\;\;\;,\;\;\;
\mu_\sigma=\mu+\frac{U}{2}(n+\sigma m),
\end{equation}
is solved self--consistently to get the ferromagnetic ground
state.
\begin{figure}
\begin{center}
\includegraphics[width=5.0cm]{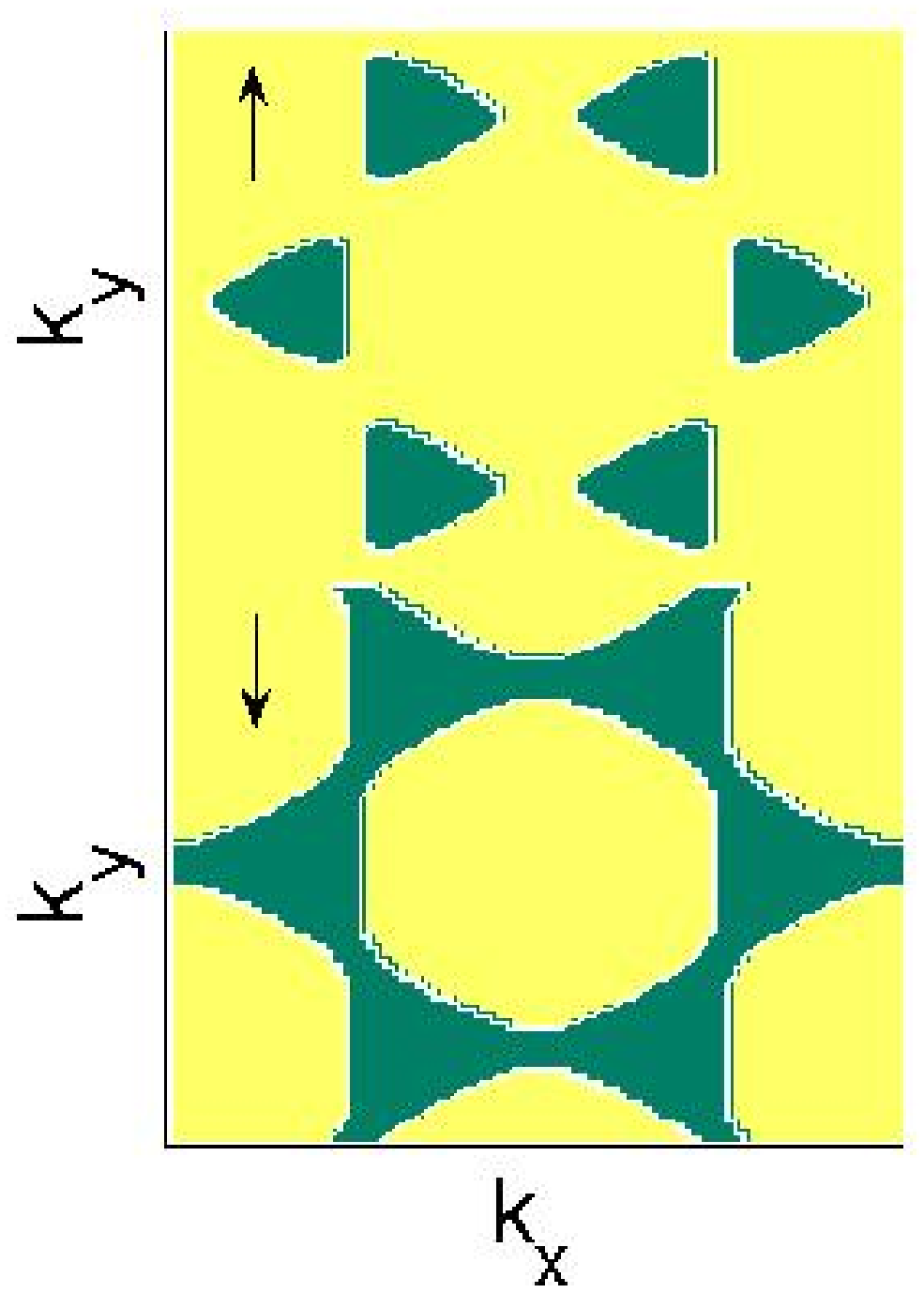}
\includegraphics[width=5.0cm]{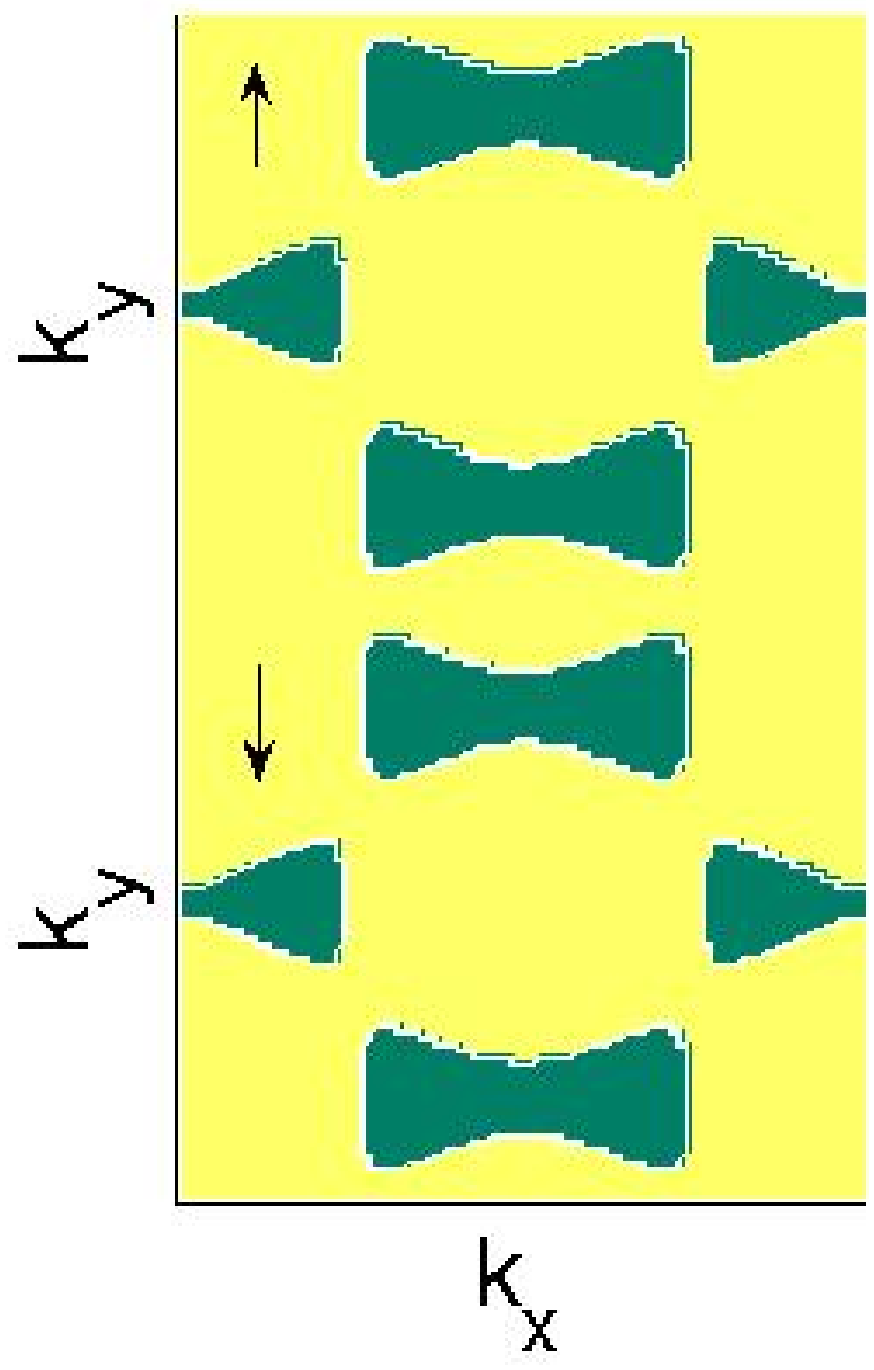}
\includegraphics[width=5.0cm]{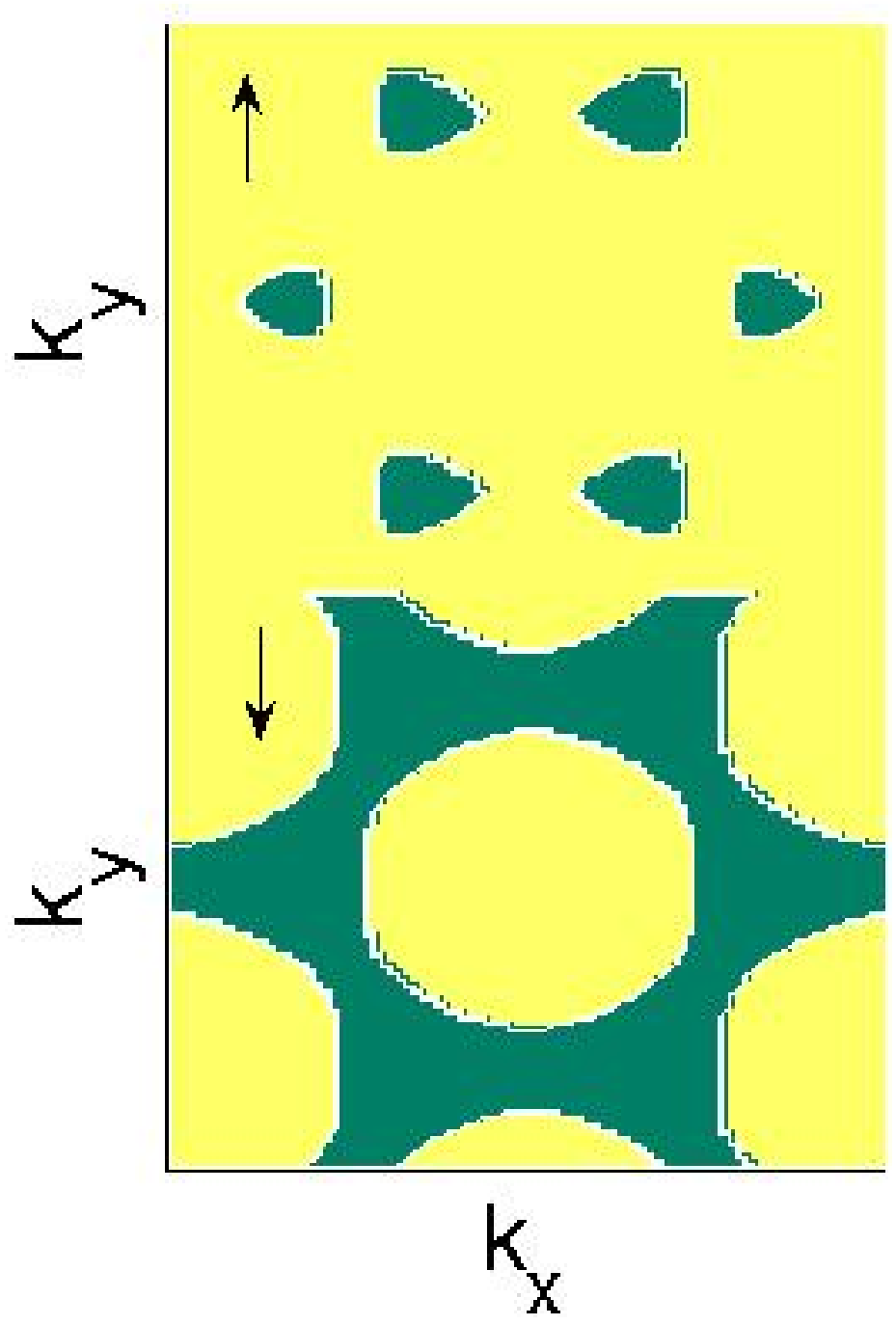}
\caption{(Color online) Left: Initial polarization of the Fermi
surface with the spin up electrons (upper part) less populated
than the spin down (lower part): $n_\downarrow-n_\uparrow=0.14$,
$\mu_\uparrow=-0.9t$, $\mu_\downarrow=-1.1t$. Center: Final state
obtained for the values of the parameters U=1t and V=2t. A
Pomeranchuk instability is clearly visible in both spin bands.
Right: Final state with enhanced ferromagnetic polarization
obtained for the values of the parameters  U=2t and V=1t.}
\label{ferro}
\end{center}
\end{figure}

The competition of the Pomeranchuk instability found in the previous
section with a possible ferromagnetic instability is studied with
the same procedure using the full mean field Hamiltonian
$H_0+H^V_{MF}+H^U_{MF}$ and solving self-consistently the equations
\begin{equation}
E_\uparrow({\bf k})=\mu+U n_\downarrow -\sqrt{t^2\vert\phi({\bf
k})\vert^2+F(V, {\bf k})_\uparrow},
\end{equation}
\begin{equation}
E_\downarrow({\bf k})=\mu+U n_\uparrow-\sqrt{t^2\vert\phi({\bf
k})\vert^2+F(V, {\bf k})_\downarrow},
\end{equation}
where $F(V, {\bf k})_\sigma$ is given in eq. (12) and
$n_\sigma=1/2(n+\sigma m)$.

\section{Summary of the results}
\label{results}

The competition of the Pomeranchuk instability with ferromagnetism
is exemplified in Fig. \ref{ferro}. The figure in the left side
represents the initial free Fermi surface where the spin up (upper
side) and down (lower side) electrons have different populations
around the Van Hove filling. In particular in the example given,
the chemical potential of the spin up (down) electrons is set
slightly below (above) the VHS: $n_\downarrow-n_\uparrow=0.14$,
$\mu_\uparrow=-0.9t$, $\mu_\downarrow=-1.1t$. The image in the
center represents the renormalized Fermi surface for the two spin
polarizations when the exchange interaction V is bigger than U:
$V=2U=2t$. We can see that the ferromagnetism has disappeared: the
final Fermi surface is the same for the  two spin polarization and
presents a Pomeranchuk deformation. The opposite case is shown in
the figure at the right: with the same initial configuration the
final state for the values of the interactions $U=2V=2t$ has an
enhanced ferromagnetism with no signal of deformation.

\begin{figure}
\begin{center}
\includegraphics[width=8cm]{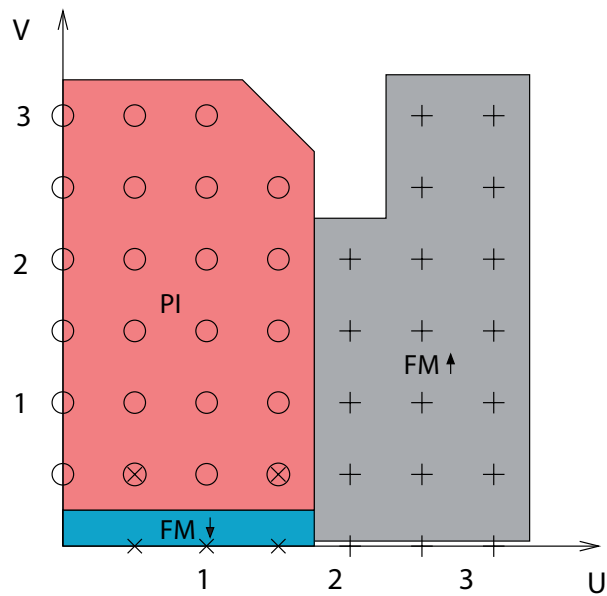}
\caption{Phase diagram showing the competition of Pomeranchuk and
ferromagnetic instabilities as a function of the interactions U
and V measured in units of the hopping parameter t.}
\label{phased2}
\end{center}
\end{figure}

The phase diagram of the dominant instability as a function of the
interaction strength U and V (measured in units of the hopping
parameter t) is shown  in Fig. \ref{phased2}. In all cases the
initial Fermi surface consists of a slightly polarized state
around the Van Hove singularity as the one shown in the left hand
side of Fig. \ref {ferro}. The symbols in Fig. \ref{phased2}
denote calculated points with the following meaning: the circles
appearing for low values of U  denote an unpolarized  final state
with Pomeranchuk deformation as the one shown in the center of
Fig. \ref{ferro}. Crosses appearing for large values of U
represent a final state where the polarization is bigger than the
initial one and the Fermi surface has the original symmetry as the
one in the right hand side of Fig. \ref{ferro}. The points denoted
by $\bigotimes$ correspond to values where both instabilities
coexist and the final Fermi surface is deformed and spin
polarized. Finally the crosses at low values of U with V=0
represent final states which are still polarized but where the
spin polarization is smaller than the initial one. This region is
denoted by $FM\downarrow$ in the figure phase diagram. PI (FM)
denotes the region where the  Pomeranchuk (ferromagnetic)
instability dominated respectively.

The ferromagnetic evolution of the system for V=0 is as follows:
Below a critical value of $1.5 < U_c < 2$ a free slightly
polarized system at fillings near (but not at) the VHS evolves
towards an unpolarized one in agreement with \cite{PAB04,PAB06}.
The Coulomb exchange V induces a deformation of the Fermi surface
already at very small values $V_c\sim 0.4 t$ when the free initial
state is near the Van Hove singularity. Increasing U enhances the
magnetic polarization of the interacting system until the critical
value $U\sim 1.75 t$ is reached where the final state corresponds
to a ferromagnetic system with un-deformed Fermi surface. Unlike
what happens in the square lattice, the critical value of U above
which ferromagnetism prevails does not depend on V (vertical line
in Fig. \ref{phased2}). In the blank region in the upper part of
the figure corresponding to high values of V around the critical U
we have not been able to reach a self--consistent solution.

We note that at V=0, the critical U for ferromagnetism is zero at
the Van Hove filling and changes very rapidly around it in a rigid
band model \cite{PAB04,PAB06}. This behavior is due to the
divergent density of states at the VHS that would be smeared by
temperature effects or disorder in real samples. Based on the
previous studies if the square lattice we are confident that the
results obtained in the present work are robust and the phase
diagram shown in fig. \ref{phased2} will remain qualitatively when
more detailed calculations are done.

The realization of one or another phase in real graphene samples
depends on the values that the effective Coulomb interaction
parametrized by $U, V$ has at the VHS filling. Besides the ARPES
experiments \cite{MBetal07} which do not comment on the strength
of the interaction, we are not aware of other experiments at these
doping levels. Conservative estimates can be obtained from the
graphite intercalated compounds \cite{TH92} on the basis of the
similarities found in  \cite{MBetal07} between the  Fermi surfaces
of the two. Values of $U\sim 2.5 t$, $V\sim t$ can be very
reasonable and lie in the range discussed in the present work.

\section{Discussion and open problems}
\label{future}

In this work we have examined some  physical issues to be expected
in graphene at high doping. We have seen that at the Van Hove
filling in the presence of an exchange Coulomb interaction the
Fermi surface is {\it softer} and becomes very easily deformed as
compared with a similar analysis of the square lattice.

A natural extension of the work presented in this article is to
study the competition of the instabilities studied in this work
with superconductivity  along the lines of ref. \cite{YM07}. A
complete renormalization group analysis of the competing low
energy instabilities of the Van Hove filling as a function of the
doping and the couplings U and V is also to be done. Due to the
special geometry of the Fermi surface in the honeycomb lattice the
symmetry breaking phases can present a richer  variety than those
of the square lattice. A fairly complete analysis of the short
range interactions around the Dirac point was done in the early
paper \cite{GGV01}. Due to the existence of two Fermi points the
classification of the possible low--energy couplings is similar to
the g--ology of the one dimensional models. The RG classification
of the low--energy couplings in the Van Hove filling is richer
since there are three independent VH points to consider and work
in this direction is in progress. An analysis of the physics of
the Fermi surface around the trigonal warping can lead also to
very interesting results in the light of the evolution of the
Fermi surface anisotropies done in \cite{GGV97,RLG06}.

On the view of the analysis of this problem done in the square
lattice we can expect that RG calculation will enrich the phase
diagram but the two phases discussed here will stay.
Ferromagnetism is a likely possibility due to the Stoner criterium
although it will compete with antiferromagnetism at high values of
U. A very important parameter to this problem is the next to
nearest neighbor hopping $t'$ that suppresses the nesting of the
bare Fermi surface and affects the shape of the phase transition
lines \cite{PAB04,PAB06}. These references focuss on the magnetism
of the Hubbard model (U) on the honeycomb lattice at finite
dopings and found that non--homogeneous (spiral) phases are the
most stable configurations around the Van Hove filling. It would
be interesting to analyze the competition and stability of these
non--homogeneous ferromagnetic configurations  in the presence of
a V interaction.

Of particular interest is the possibility of coexistence or
competition of the Pomeranchuk instability with possible
superconducting instabilities \cite{G08}. A very interesting
suggestion has been made in studies of the square lattice that
superconductivity changes the nature of the Pomeranchuk transition
going from first to second order \cite{YM07}. This feature will
probably be maintained in the honeycomb lattice and we are
actually exploring this problem.


 As it is known, the Kohn--Luttinger instability in its original context
\cite{KL65} suggests that no Fermi liquid will be stable at
sufficiently low temperatures. In the case of a 3D electron system
with isotropic Fermi surface there is an enhanced scattering at
momentum transfer $2k_F$  which translates into a modulation of
the effective interaction potential with oscillating behavior
similar to the Friedel oscillations. This makes possible the
existence of attractive channels, labelled by the angular momentum
quantum number.  The issue of the stability of Fermi liquids was
studied rigorously in two dimensions in a set of papers
\cite{FKT04a,FKT04b,FKT04c} with the conclusion that a Fermi
liquid could exist in two space dimensions at zero temperature
provided that the Fermi surface of the system obeys some
"asymmetry" conditions. In particular the non--interacting Fermi
surface had to lack inversion symmetry
$$\epsilon(-{\bf k})\neq\epsilon({\bf k})$$
at all points k and be otherwise quite regular (the Van Hove
singularities can not be present).  The dispersion relation of
graphene obeys such a condition for fillings where the trigonal
warping is already noticeable and below the Van Hove filling
provided that disorder and interactions do not mix the Van Hove
points. This is a very common assumption in the graphene physics
around the Dirac points where the inversion symmetry in k--space
acts as a time reversal symmetry \cite{MGV07}. We find interesting
to note that the --quite restrictive--conditions of refs.
\cite{FKT04a,FKT04b,FKT04c} can be fulfilled in a real --and very
popular--system.

\section{Acknowledgments.}

We thank A. Cortijo, M. P. L\'opez-Sancho and R. Rold\'an for very
interesting discussions. This research was supported by the
Spanish MECD grant FIS2005-05478-C02-01 and by the {\it
Ferrocarbon} project from the European Union under Contract 12881
(NEST).
\bibliography{VH}
\end{document}